# Generalized Formulation of Weighted Optimal Guidance Laws with Impact Angle Constraint

Chang-Hun Lee[1], Min-Jea Tahk[2], and Jin-Ik Lee[3]
*Korea Advanced Institute of Science and Technology (KAIST), Daejeon, 305-701, Korea*

**The purpose of this paper is to investigate the generalized formulation of weighted optimal guidance laws with impact angle constraint. From the generalized formulation, we explicitly find the feasible set of weighting functions that lead to analytical forms of weighted optimal guidance laws. This result has potential significance because it can provide additional degrees of freedom in designing a guidance law that accomplishes the specified guidance objective.**

## Notice

[1] Ph. D student, Department of Aerospace Engineering, Korea Advanced Institute of Science and Technology(KAIST), Kuseong Yuseong, Daejeon, 305-701, Korea/chlee@fdcl.kaist.ac.kr.
[2] Professor, Department of Aerospace Engineering, Korea Advanced Institute of Science and Technology(KAIST), Kuseong Yuseong, Daejeon, 305-701, Korea/mjtahk@fdcl.kaist.ac.kr
[3] Principal researcher, Agency for Defense Development, PO Box 35-3, Yuseong, Daejeon, 305-600, Korea/jilee@fdcl.kaist.ac.kr

# I. Introduction

For anti-ship and anti-tank missile systems, the guidance laws achieving the desired impact angle have been considered in order to maximize the warhead effect and attack a target's weak spot. Over the past decades, the optimal control theory [1] has been extensively used to design the impact angle control guidance laws because of its benefits: It can easily provide a guidance law that satisfies the terminal constraints and some performance requirements as well as the analytical form and state feedback form of guidance laws. In the application of optimal control for deriving the impact angle control guidance laws, the minimization of the control effort has been widely considered for the cost function [2-8] as follows:

$$J = \int_{t_0}^{t_f} u^2(\tau) d\tau \tag{1}$$

where $u$ and $t_f$ represent the missile's acceleration command and the time of interception, respectively.

In this method, how to design the cost function is an important issue for guidance law designers because the selection of the cost function can affect the response of state variables and then decide the guidance performance. Accordingly, to improve the guidance performance, the control energy costs with weighting functions have also been used to derive the impact angle control guidance laws. A power of time-to-go function [9-11] and an exponential function [12-14] were considered for the weighting function of the energy cost. In such previous works, for achieving the specified guidance purpose, the weighted cost functions were introduced to shape the missile's trajectory or to distribute the acceleration demand during the engagement.

These previous studies now raise a question: Could any weighting function be used for accomplishing such guidance objectives? The purpose of this paper is to find answers to this question. In this paper, we first investigate the generalized formulation of the weighted optimal control problems with the terminal constraints (i.e., zero miss distance and the desired impact angle) as follows:

$$J = \int_{t_0}^{t_f} W(\tau) u^2(\tau) d\tau \tag{2}$$

Then, we determine the feasible set of $W(t)$ that lead to the analytical solutions because the analytical forms of solutions are more desirable for practical uses. From a practical standpoint, the potential significance of this result is that through appropriate selections in the set of the weighting functions we have determined, the designer can achieve the guidance purpose as desired.

## II. Problem Formulation

Let us consider the engagement geometry for a stationary target as described in Fig. 1, where $(X_I, Y_I)$ and $(x_f, y_f)$ denote the inertial reference frame and the impact angle frame, respectively. The impact angle frame is defined to be rotated from the inertia reference frame by $\gamma_f$, which is the desired impact angle. The flight path angle and the line-of-sight angle are denoted by $\gamma_M$ and $\sigma$. In the impact angle frame, these angles are expressed as follows:

$$\bar{\gamma}_M = \gamma_M - \gamma_f, \qquad \bar{\sigma} = \sigma - \gamma_f \tag{3}$$

In Fig. 1, other variables are self-explanatory, and the engagement kinematics with respect to the impact angle frame can be written as:

$$\begin{aligned} \dot{y} &= V_M \sin \bar{\gamma}_M \\ \dot{\bar{\gamma}}_M &= a_M / V_M \end{aligned} \tag{4}$$

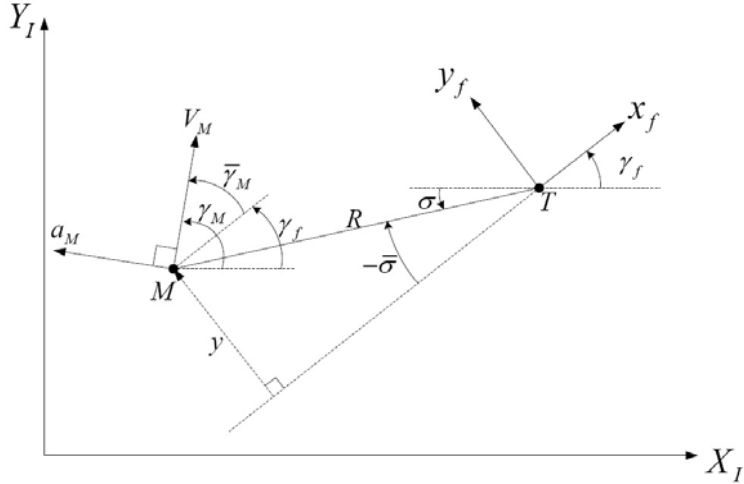

Fig. 1 The homing engagement geometry and parameter definitions.

It is assumed that $V_M$ is constant and $\bar{\gamma}_M$ is small enough to linearize the engagement kinematics as follows:

$$\begin{aligned} \dot{y} &= V_M \bar{\gamma}_M = v \\ \dot{v} &= a_M \end{aligned} \tag{5}$$

where $y$ and $v$ represent the lateral distance and velocity perpendicular to the desired impact course. In the linearized engagement kinematics, the flight path angle and LOS angle with respect to the impact angle frame can be determined, respectively, as follows:

$$\bar{\gamma}_M = \frac{v}{V_M}, \qquad \bar{\sigma} = -\frac{y}{R} = -\frac{y}{V_M t_{go}} \tag{6}$$

By using Eqs. (3) and (6), the lateral distance and velocity perpendicular to the impact course can be rewritten in terms of the flight path angle and LOS angle, respectively.

$$y = V_M t_{go} (\gamma_f - \sigma), \qquad v = V_M (\gamma_M - \gamma_f) \tag{7}$$

The linearized engagement kinematics as given in Eq. (5) can be rewritten in the matrix form as:

$$\dot{x} = Ax + Bu \tag{8}$$

where,

$$x \triangleq [y, v]^T, \qquad u \triangleq a_M, \qquad A \triangleq \begin{bmatrix} 0 & 1 \\ 0 & 0 \end{bmatrix}, \qquad B \triangleq \begin{bmatrix} 0 \\ 1 \end{bmatrix} \tag{9}$$

In order to satisfy the zero miss distance and the desired impact angle at the terminal time, the following boundary conditions should be achieved.

$$x_1(t_f) = x_2(t_f) = 0 \tag{10}$$

Now, let us set the following optimal control problem which minimizes the control effort weighted by general functions of $W(t)$.

$$\min_{u} \; J = \frac{1}{2} \int_{t_0}^{t_f} W(\tau) u^2(\tau) d\tau, \qquad \text{where,} \quad W(\tau) > 0 \; \text{for} \; \tau \in (t_0, t_f) \tag{11}$$

The conditions of feasible weighting functions will be determined in the next section.

### III. Generalized Formulation of Weighted Optimal Solutions

In this paper, we use Schwarz's Inequality approach as studied in [2] in order to solve the optimal problem. First, according to the linear control theory, the general solution of Eq. (8) can be expressed as:

$$x(t_f) = \Phi(t_f - t) x(t) + \int_{t}^{t_f} \Phi(t_f - \tau) B(\tau) u(\tau) d\tau \tag{12}$$

where $\Phi(t_f - t)$ represents the state transition matrix and is determined.

$$\Phi(t_f - t) \triangleq e^{A(t_f - t)} = \begin{bmatrix} 1 & t_f - t \\ 0 & 1 \end{bmatrix} \tag{13}$$

Then, Eq. (12) provides the expression of state variable at the final time.

$$x_1(t_f) = f_1 - \int_t^{t_f} h_1(\tau)u(\tau)d\tau \tag{14}$$

$$x_2(t_f) = f_2 - \int_t^{t_f} h_2(\tau)u(\tau)d\tau \tag{15}$$

where,

$$\begin{aligned} f_1 &\triangleq x_1(t) + (t_f - t)x_2(t), & h_1(\tau) &\triangleq -(t_f - \tau) \\ f_2 &\triangleq x_2(t), & h_2(\tau) &\triangleq -1 \end{aligned} \tag{16}$$

By imposing the boundary conditions (i.e., $x_1(t_f) = x_2(t_f) = 0$), Eqs. (14) and (15) can be rewritten as:

$$f_1 = \int_t^{t_f} h_1(\tau)u(\tau)d\tau \tag{17}$$

$$f_2 = \int_t^{t_f} h_2(\tau)u(\tau)d\tau \tag{18}$$

Hereafter, let us introduce a new variable denoted by $\lambda$. Then, we can combine Eqs. (17) and (18) as follows:

$$f_1 - \lambda f_2 = \int_t^{t_f} [h_1(\tau) - \lambda h_2(\tau)]u(\tau)d\tau \tag{19}$$

The above equation can be rewritten by introducing a slack variable with respect to $W(\tau)$.

$$f_1 - \lambda f_2 = \int_t^{t_f} [h_1(\tau) - \lambda h_2(\tau)]W^{-1/2}(\tau)W^{1/2}(\tau)u(\tau)d\tau \tag{20}$$

Then, applying Schwarz's Inequality to Eq. (20) and rearranging the obtained result yields the following inequality condition:

$$\frac{(f_1 - \lambda f_2)^2}{2\int_t^{t_f} [h_1(\tau) - \lambda h_2(\tau)]^2 W^{-1}(\tau)d\tau} \leq \frac{1}{2}\int_t^{t_f} W(\tau)u^2(\tau)d\tau \tag{21}$$

Note that the right hand side of Eq. (21) is equal to the cost function defined in Eq. (11). It can predict that when the equality (i.e., =) holds, the left hand side is identical to the minimum value of the cost function. According to Schwarz's Inequality, the acceleration command that holds the equality can be expressed as:

$$u(\tau) = K[h_1(\tau) - \lambda h_2(\tau)]W^{-1}(\tau) \tag{22}$$

where $K$ is a constant to be determined. Eq. (22) can be regarded as the acceleration command that minimizes the cost function. Substituting Eq. (22) into Eq. (17) gives the following equation.

$$K = \frac{f_1}{\int_t^{t_f} h_1^2(\tau)W^{-1}(\tau)d\tau - \lambda \int_t^{t_f} h_1(\tau)h_2(\tau)W^{-1}(\tau)d\tau} \tag{23}$$

For convenience, we introduce shorthand notations as follows:

$$g_1 \triangleq \int_t^{t_f} h_1^2(\tau) W^{-1}(\tau) d\tau \tag{24}$$

$$g_{12} \triangleq \int_t^{t_f} h_1(\tau) h_2(\tau) W^{-1}(\tau) d\tau \tag{25}$$

$$g_2 \triangleq \int_t^{t_f} h_2^2(\tau) W^{-1}(\tau) d\tau \tag{26}$$

Using these notations provides a simplified expression of $K$ as:

$$K = \frac{f_1}{g_1 - \lambda g_{12}} \tag{27}$$

Substituting Eq. (27) into Eq. (22), we have the following equation.

$$u(\tau) = \frac{f_1 [h_1(\tau) - \lambda h_2(\tau)] W^{-1}(\tau)}{g_1 - \lambda g_{12}} \tag{28}$$

From Eq. (21), the minimum value of the cost function can be expressed using shorthand notations:

$$J = \frac{(f_1 - \lambda f_2)^2}{2(g_1 - 2\lambda g_{12} + \lambda^2 g_2)} \tag{29}$$

Because the undetermined value $\lambda$ exists in Eq. (29), this expression is incomplete. From the calculus, we can find $\lambda$ which further minimizes $J$ by taking the derivative of $J$ with respect to $\lambda$ and then by imposing $dJ/d\lambda = 0$ as follows:

$$\lambda^* = \frac{f_1 g_{12} - f_2 g_1}{f_1 g_2 - f_2 g_{12}} \tag{30}$$

Substituting Eq. (30) into Eq. (28), we have the optimal acceleration command.

$$u^*(\tau) = \frac{\left[ f_1 h_1(\tau) g_2 - g_{12} (f_1 h_1(\tau) + f_1 h_2(\tau)) + f_2 h_2(\tau) g_1 \right] W^{-1}(\tau)}{g_1 g_2 - g_{12}^2} \tag{31}$$

In the time domain, the optimal acceleration command can be further simplified by substituting Eqs. (9) and (16) into Eq. (31) and by introducing newly defined variables called the equivalent guidance gains.

$$a_M^* = -k_1 \frac{y}{t_{go}^2} - k_2 \frac{v}{t_{go}} \tag{32}$$

where $t_{go} \triangleq t_f - t$ is the remaining time of interception. The notations $k_1$ and $k_2$ represent the equivalent guidance gains and are defined as follows:

$$k_1 = \left( \frac{g_2 t_{go}^3 - g_{12} t_{go}^2}{g_1 g_2 - g_{12}^2} \right) W^{-1}(t) \tag{33}$$

$$k_2 = \left( \frac{g_1 t_{go} + g_2 t_{go}^3 - 2g_{12} t_{go}^2}{g_1 g_2 - g_{12}^2} \right) W^{-1}(t) \tag{34}$$

where

$$g_1 = \int_t^{t_f} (t_f - \tau)^2 W^{-1}(\tau) d\tau \tag{35}$$

$$g_{12} = \int_t^{t_f} (t_f - \tau) W^{-1}(\tau) d\tau \tag{36}$$

$$g_2 = \int_t^{t_f} W^{-1}(\tau) d\tau \tag{37}$$

From Eqs. (33) and (34), the denominator of $k_1$ and $k_2$ should not be zero. It can be proven by the following lemma.

**Lemma 1.** Regardless of choices in the weighting functions, the following relation is always guaranteed.

$$g_1 g_2 - g_{12}^2 > 0 \tag{38}$$

*Proof.* From Eq. (36), the expression of $g_{12}$ can be reformulated as:

$$g_{12} = \int_t^{t_f} (t_f - \tau) W^{-1/2}(\tau) W^{-1/2}(\tau) d\tau \tag{39}$$

Applying Schwartz's Inequality to Eq. (39) yields the following results.

$$g_{12}^2 \leq \int_t^{t_f} (t_f - \tau)^2 W^{-1}(\tau) d\tau \int_t^{t_f} W^{-1}(\tau) d\tau \tag{40}$$

The equality sign of Eq. (40) does not hold because of $(t_f - \tau) W^{-1/2}(\tau) \neq \alpha W^{-1/2}(\tau)$, where $\alpha$ is a constant. Therefore, the final result is written using the terms of $g_1$ and $g_2$ as follows:

$$g_1 g_2 - g_{12}^2 > 0 \tag{41}$$

which completes the proof. ∎

## IV. Feasible Set of Weighting Functions

The expressions in Eqs. (32) through (37) represent the generalized formulation of weighed optimal acceleration commands satisfying zero miss distance as well as the impact angle constraint. According to selections of weighting functions, the state feedback form of optimal guidance law can be determined by computing Eqs. (35) through (37) and then by substituting these results into Eqs. (33) and (34). Accordingly, in order to obtain analytical forms of weighted optimal guidance laws, we require the weighting functions as provided in the following proposition.

**Proposition 1.** If the weighting function $W(\tau)$ satisfies the condition of Eq. (11) and the integrations of $W(\tau)$ as shown in Eq. (42) are analytically given, then this weighting function can lead to the analytical solution.

$$W_0(\tau) \triangleq W^{-1}(\tau), \quad W_1(\tau) \triangleq \int W_0(\tau)d\tau, \quad W_2(\tau) \triangleq \int W_1(\tau)d\tau, \quad W_3(\tau) \triangleq \int W_2(\tau)d\tau \tag{42}$$

where $W_1(\tau)$, $W_2(\tau)$, and $W_3(\tau)$ represent the indefinite integral, double integral, and triple integral of the inversed weighting function.

*Proof.* From Eqs. (33) and (34), in order to obtain analytical solutions, the composition terms of the equivalent guidance gains should be analytically given. Therefore, the inverse of weighting function $W^{-1}(\tau)$ basically has an analytical form and the terms of $g_1$, $g_{12}$, and $g_2$ are also given in analytical forms. From the calculus, the terms of $g_1$, $g_{12}$, and $g_2$ can be further expanded based on the method of integration by parts.

$$g_1 = -t_{go}^2 W_1(t) - 2t_{go} W_2(t) + 2\left(W_3(t_f) - W_3(t)\right) \tag{43}$$

$$g_{12} = -t_{go} W_1(t) + W_2(t_f) - W_2(t) \tag{44}$$

$$g_2 = W_1(t_f) - W_1(t) \tag{45}$$

Accordingly, the analytical result of $W_1(\tau)$, $W_2(\tau)$, and $W_3(\tau)$ introduce analytical expressions of $g_1$, $g_{12}$, and $g_2$, which completes the proof. ∎

Note that the condition in Eq. (42) represents the feasible set of weighting functions to obtain the analytical weighted optimal guidance laws. Additionally, to ensure that the guidance command does not blow up during the engagement, the equivalent guidance gain should be bounded: The functions of $W^{-1}(\tau)$ is bounded as $W^{-1}(\tau) < \infty$ for $\tau \in (t_0, t_f)$.

From a practical standpoint, these results are helpful to derive a new guidance law that improves the guidance performance and attains the specific guidance objective through appropriate choices in the feasible set of weighting functions, which satisfy the previously determined conditions.

Hereafter, we illustrate our results with two simple cases as the weighting functions $W^{-1}(\tau) = 1$ and $W^{-1}(\tau) = (t_f - \tau)^N$, which comply with the conditions as we have discussed. For $W^{-1}(\tau) = 1$, which means the control effort is equally weighted during the engagement, we have the following results.

$$g_1 = \int_t^{t_f} (t_f - \tau)^2 \, d\tau = \frac{1}{3} t_{go}^3 \tag{46}$$

$$g_{12} = \int_t^{t_f} (t_f - \tau) \, d\tau = \frac{1}{2} t_{go}^2 \tag{47}$$

$$g_2 = \int_t^{t_f} 1 \, d\tau = t_{go} \tag{48}$$

We substitute Eqs. (46), (47), and (48) into Eqs. (33) and (34) under the condition of $W^{-1}(\tau) = 1$. Then, the optimal acceleration command is obtained in that case as follows:

$$a_M^* = -6 \frac{y}{t_{go}^2} - 4 \frac{v}{t_{go}} \tag{49}$$

Since the lateral distance and velocity are defined the impact angle frame, the terminal values of these variables are zero as $y_f = v_f = 0$. Rearranging Eq. (49) and using the terminal condition $y_f = v_f = 0$ yield the following alternative form of this acceleration command.

$$a_M^* = \frac{6}{t_{go}^2} \left[ y_f - y - v t_{go} \right] - \frac{2}{t_{go}} \left[ v_f - v \right] \tag{50}$$

In addition, substituting Eq. (7) into Eq. (49), the guidance command can be rewritten as follows:

$$a_M^* = -\frac{V_M}{t_{go}} \left[ -6\sigma + 4\gamma_M + 2\gamma_f \right] \tag{51}$$

Note that these results are identical to the optimal control guidance law with terminal impact angle constraint as studied in [1-3,8]. If we choose $W^{-1}(\tau) = (t_f - \tau)^N$, which increases the weight of acceleration demand as $t \to t_f$, then $g_1$, $g_{12}$, and $g_2$ can be computed as:

$$g_1 = \int_t^{t_f} (t_f - \tau)^{N+2} \, d\tau = \frac{1}{N+3} t_{go}^{N+3} \tag{52}$$

$$g_{12} = \int_t^{t_f} (t_f - \tau)^{N+1} \, d\tau = \frac{1}{N+2} t_{go}^{N+2} \tag{53}$$

$$g_2 = \int_t^{t_f} (t_f - t)^N \, d\tau = \frac{1}{N+1} t_{go}^{N+1} \tag{54}$$

Then, we have the following optimal acceleration command.

$$a_M^* = -(N+3)(N+2) \frac{y}{t_{go}^2} - 2(N+2) \frac{v}{t_{go}} \tag{55}$$

In a similar way, alternative forms of this command can be obtained as:

$$a_M^* = \frac{(N+3)(N+2)}{t_{go}^2}\left[y_f - y - vt_{go}\right] - \frac{(N+1)(N+2)}{t_{go}}\left[v_f - v\right] \tag{56}$$

$$a_M^* = -\frac{V_M}{t_{go}}\left[-(N+3)(N+2)\sigma + 2(N+2)\gamma_M + (N+1)(N+2)\gamma_f\right] \tag{57}$$

where $N$ is the power of time-to-go. Note that these results are equal to the time-to-go weighted optimal control guidance laws as studied in [9-10].

## V. Conclusion

In this paper, optimal guidance laws with terminal impact angle constraint are generalized for the weighted control energy costs. The results indicated that any weighting function can provide the analytical form of optimal solution if up to triple integrations of the inverse of the weighting functions are analytically given. The potential significance of these results is that the feasible set of weighting functions as we have determined can provide additional degrees of freedom for designing a guidance law that achieves the guidance purpose as desired and enhances the guidance performance.

## Acknowledgments

This research was supported by Agency for Defense Development under the contract UD110031CD.